\documentclass[reprint,prx,amsmath,amssymb,superscriptaddress,floatfix,longbibliography]{revtex4-1}
\usepackage{hyperref}
\usepackage{graphics}
\usepackage{enumerate}
\usepackage{color}
\usepackage{mathrsfs}
\usepackage[normalem]{ulem}

\RequirePackage{lineno} 

\newcommand{\ket}[1]{\left|{#1}\right\rangle}
\newcommand{\bk}[1]{\left\langle{#1}\right\rangle}

\begin{document}

\title{Valley filtering in spatial maps of coupling between silicon donors and quantum dots}

\author{J. Salfi}
\affiliation{Centre for Quantum Computation and Communication Technology, School of Physics, The University of New South Wales, Sydney, NSW 2052, Australia.}

\author{B. Voisin}
\affiliation{Centre for Quantum Computation and Communication Technology, School of Physics, The University of New South Wales, Sydney, NSW 2052, Australia.}

\author{A. Tankasala}
\affiliation{Purdue University, West Lafayette, IN 47906, USA.}

\author{J. Bocquel}
\affiliation{Centre for Quantum Computation and Communication Technology, School of Physics, The University of New South Wales, Sydney, NSW 2052, Australia.}

\author{M. Usman}
\affiliation{Centre for Quantum Computation and Communication Technology, School of Physics, University of Melbourne, Parkville, VIC 3010, Australia.}

\author{M. Y. Simmons}
\affiliation{Centre for Quantum Computation and Communication Technology, School of Physics, The University of New South Wales, Sydney, NSW 2052, Australia.}

\author{L. C. L. Hollenberg}
\affiliation{Centre for Quantum Computation and Communication Technology, School of Physics, University of Melbourne, Parkville, VIC 3010, Australia.}

\author{R. Rahman}
\affiliation{Purdue University, West Lafayette, IN 47906, USA.}

\author{S. Rogge}
\affiliation{Centre for Quantum Computation and Communication Technology, School of Physics, The University of New South Wales, Sydney, NSW 2052, Australia.}

\begin{abstract}
Exchange coupling is a key ingredient for spin-based quantum technologies since it can be used to entangle spin qubits and create logical spin qubits. However, the influence of the electronic valley degree of freedom in silicon on exchange interactions is presently the subject of important open questions. Here we investigate the influence of valleys on exchange in a coupled donor/quantum dot system, a basic building block of recently proposed schemes for robust quantum information processing.  Using a scanning tunneling microscope tip to position the quantum dot with sub-nm precision, we find a near monotonic exchange characteristic where lattice-aperiodic modulations associated with valley degrees of freedom comprise less than 2~\% of exchange. From this we conclude that intravalley tunneling processes that preserve the donor's $\pm x$ and $\pm y$ valley index are filtered out of the interaction with the $\pm z$ valley quantum dot, and that the $\pm x$ and $\pm y$ intervalley processes where the electron valley index changes are weak.  Complemented by tight-binding calculations of exchange versus donor depth, the demonstrated electrostatic tunability of donor/QD exchange can be used to compensate the remaining intravalley $\pm z$ oscillations to realise uniform interactions in an array of highly coherent donor spins. 
\end{abstract}

\date{\today}
\maketitle
Following proposals for spin-based quantum computing\cite{Kane:1998ce,Loss:1998ia}, spin qubits have been demonstrated in, \textit{e.g.}, diamond\cite{Childress:2006km}, GaAs\cite{Petta:2005kn,Nowack:2011cu,Shulman:2012fk,Medford:2013bl}, Si donors\cite{Pla:2012jj} and Si quantum dots (QDs)\cite{Maune:2012iu,Kim:2014dj,Veldhorst:2015jea,Reed:2016dj}. Exchange coupling plays a key role in these proposals\cite{Kane:1998ce,Loss:1998ia} and has been employed experimentally to couple spins over short distances\cite{Nowack:2011cu,Veldhorst:2015jea}, and to define multi-spin qubits\cite{Petta:2005kn,Maune:2012iu,Shulman:2012fk,Medford:2013bl,Kim:2014dj,Reed:2016dj} that can be coupled over larger distances via electric interactions\cite{Shulman:2012fk}, as also expected for spin-orbit qubits\cite{Flindt:2006dn,Nowack:2007du,NadjPerge:2010kw,Salfi:2016ct,Salfi:2016dm}. Because of the importance of exchange interactions, the impact of silicon's valley degrees freedom on electron tunneling and exchange has been the subject of many theoretical studies\cite{Cullis:1970jp,Koiller:2001gwa,Koiller:2002ih,Wellard:2005hs,Testolin:2007ku,Hill:2007ku,Pica:2014hr,Gamble:2015cl,Wang:2016ha,Nielsen:2012wt,Zimmerman:2017hg}.  Notably, small changes in donor position\cite{Cullis:1970jp,Koiller:2001gwa,Koiller:2002ih,Wellard:2005hs,Pica:2014hr,Gamble:2015cl,Wang:2016ha} and QD surface roughness\cite{Nielsen:2012wt,Zimmerman:2017hg} are expected to produce large modulations of exchange coupling, affecting two-qubit gate fidelities, owing to the lattice-aperiodicity of the valley wavevector.  For donors, the negative effects of the predicted rapid non-monotonic dependence of exchange could be be reduced by atomic precision placement\cite{Fuechsle:2012bl,Weber:2014ez} and/or quantum control schemes to recover two-qubit gate infidelities\cite{Testolin:2007ku,Hill:2007ku}. While predictions of the amplitude of the non-monotonic oscillations vary significantly\cite{Cullis:1970jp,Koiller:2001gwa,Koiller:2002ih,Wellard:2005hs,Testolin:2007ku,Hill:2007ku,Pica:2014hr,Gamble:2015cl,Wang:2016ha,Nielsen:2012wt,Zimmerman:2017hg}, experimentally establishing the strength of the exchange modulations has proven a difficult task and experimentally probing the role valleys in exchange has received no direct attention. 

Tunneling and exchange in coupled donor/QD systems\cite{Lansbergen:2008bs,Foote:2015ca,Urdampilleta:2015bn,HarveyCollard:2017ic} underpin some recent theory proposals for robust spin-based quantum computing\cite{Srinivasa:2015bm,Pica:2016bb,Tosi:2017fs} seeking to exploit the long donor spin coherence times\cite{Tyryshkin:2011fi,Steger:2012ev,Wolfowicz:2013ix} without direct exchange between donors.  The role of valleys in coupled donor/QD systems differs compared to the more well studied case of two donors\cite{Dehollain:2014kf,GonzalezZalba:2014kp,Weber:2014ez}: the absence of $\pm x$ and $\pm y$ valleys in the two-valley ($\pm z$) QD state means that intravalley exchange processes, where electrons preserve their valley index, occur for the $\pm z$ valleys but not for $\pm x$ and $\pm y$ valleys of the donor. Though not yet observed experimentally, this {\it filtering} of $\pm x$ and $\pm y$ valley degrees of freedom from intravalley donor/QD exchange should eliminate the main source rapid non-monotonic variations of exchange with in-plane donor position.  However, weaker intervalley processes where electrons change their valley index\cite{Ning:1971kg,Pantelides:1974ewa} remain a potential source of rapid non-monotonic exchange variations.  Ignored in exchange calculations to date\cite{Cullis:1970jp,Koiller:2001gwa,Koiller:2002ih,Wellard:2005hs,Testolin:2007ku,Hill:2007ku,Pica:2014hr,Gamble:2015cl}, intervalley processes become stronger as wavefunctions gets smaller\cite{Ning:1971kg,Pantelides:1974ewa}. Large variations in exchange are also expected with donor depth variations, due to $\pm z$ intravalley tunneling, where the phase of the donor (QD) electron is pinned by the ion (interface). The extent to which these processes influence schemes for donor/QD based quantum computing has yet to be established\cite{Srinivasa:2015bm,Pica:2016bb,Tosi:2017fs}. 

\begin{figure*}
\includegraphics{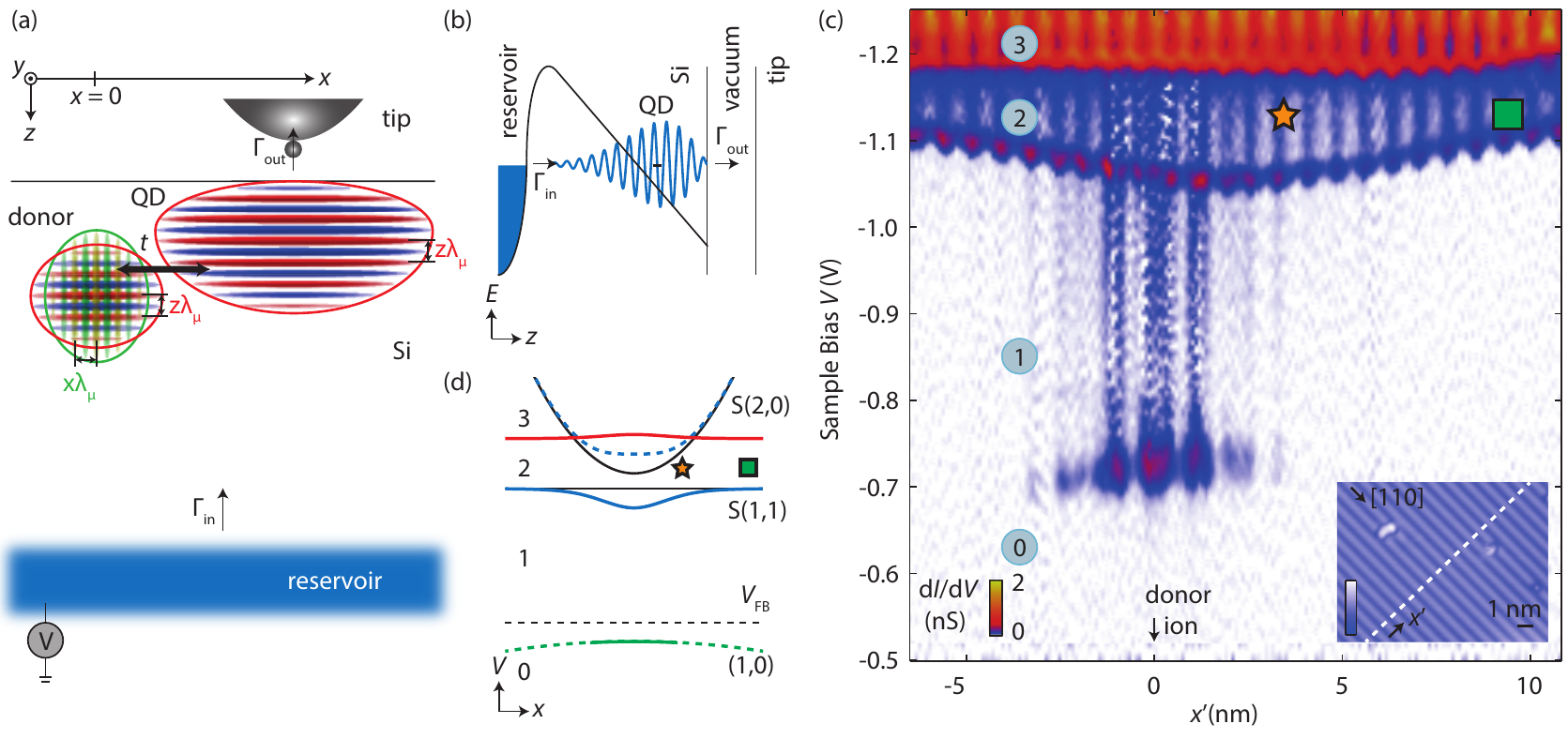}
\caption{Schematic of single-electron tunnelling from buried reservoir, to coupled donor/QD state, to the tip. The donor wavefunction parameters are taken from ref.~\onlinecite{Saraiva:2016bs}, and the $\pm x$ and $\pm z$ valleys participating in the six valley superposition are labeled.  The $\pm y$ valleys of the donor are omitted for clarity.  $\lambda_\mu=2\pi/k_\mu$, where $k_\mu=0.82(2\pi/a_0)$ is the conduction band minimum wavevector, and $a_0$ is silicon's lattice constant.  (b) Schematic diagram of isolated QD resonance created by tip-induced band bending. (c) Measured d$I$/d$V$ spectrum labelling the 0,1,2 (donor/QD hybrid) and 3 electron regions along the $110$ crystal direction. Inset: measured topography where dashed line shows locations for d$I$/d$V$ data. (d) Energy diagram for transitions between 0, 1, 2, and 3 electron states labeled with green, blue, and red lines, respectively.  Tunnel coupling $t$ between the donor and QD hybridizes the two-electron states where 2 electrons are on the donor and where there is 1 electron each on the donor and QD (blue lines).  Unhybridized 2 electron states are shown as black lines and hybridized states as blue lines. The star (box) denotes the QD position for strong (weak) hybridization of two-electron states. Temperature: 4.2 K.}
\label{fschem}
\end{figure*}

Here we experimentally investigate whether changes in lateral QD position can overcome variations of donor/QD exchange associated with high spatial frequency ``valley" oscillations and slow envelope function decay, for three dimensional donor positioning uncertainty.  This is accomplished by experimentally probing the exchange coupling $J$ of a donor bound electron with a highly localized electrostatic QD whose lateral position $\bf{R}$ relative to the donor can be controlled with sub-nm precision. The single-electron QD, which has a large $\sim12$ meV charging energy and correspondingly few-nm small spatial extent, is formed beneath a passivated Si surface and its position, and coupling to donors, is controlled by moving the tip laterally above the sample surface.  Here, we measure the donor/QD energy spectrum by single-electron transport\cite{Salfi:2014kaa,Voisin:2015gl} to quantify the strength of the intervalley interference processes in the exchange coupling $J(\bf{R})$. We note that the tunability of the exchange interaction opens up interesting possibilities to electrically probe small-scale dopant-based quantum simulators\cite{Salfi:2016cm,Le:2017jt,Dusko:2018gv}, or to perform electrical spin readout on optically active impurity centers in materials like silicon carbide\cite{Koehl:2011fv,Castelletto:2013dg}, silicon\cite{Buckley:2017hu,Beaufils:2018fa}, and diamond\cite{Brenneis:2015dq,Bourgeois:2015ke}.

We find that lattice-aperiodic exchange has a small amplitude $\lesssim2\%$ of the nominal exchange coupling, evidencing the valley filtering effect and setting an upper bound on the intervalley tunneling strength.  We also experimentally explore the tunability of the donor/QD exchange with $\bf{R}$, and find that a modest $6$ nm lateral QD shift changes the donor/QD exchange by an order of magnitude. Finally, we show that the QD also has a negligible impact on the electronic orbital and valley population of the donor, which is $3.6$ nm beneath the Si surface, which is important for some proposals\cite{Srinivasa:2015bm,Pica:2016bb}.  Using sp$^3$d$^5$s$^\ast$ calculations, we find that the observed tunability of donor/QD exchange can readily compensate variations in exchange due to nm scale donor depth uncertainty. These results show that valley-induced variations in donor/QD exchange can be (i) altogether neglected for in-plane donor positioning variations due to the valley filtering effect and weak intervalley scattering, and (ii) compensated for donor depth variations by modest electrostatic tuning of QD wavefunctions using surface gates\cite{Srinivasa:2015bm,Pica:2016bb,Tosi:2017fs}, without distorting the donor wavefunction. 

The key ingredient in our experiment is a single-electron QD whose coupling to individual donors can be tuned by controlling the QD position with sub-nm precision, using a scanning tunnelling microscope (STM) tip (Fig.~\ref{fschem}a).  The QD state is formed below a silicon/vacuum surface when the bands are locally bent downwards by the tip due to a bias $V$ applied to a reservoir (Fig.~\ref{fschem}b)\cite{Morgenstern:2001jx,Freitag:2016bl}.  The QD and donor are contained in a lightly doped region, above a highly doped reservoir and below a (100) hydrogen terminated surface (Fig.~\ref{fschem}a).  The doping gradient was prepared by thermal annealing\cite{Salfi:2014kaa}.

The energy of the QD was probed by spatially resolved single-electron tunneling. For the data shown in the ${\rm d} I/ {\rm d} V$ map of Fig.~\ref{fschem}c containing a neutral donor resonance\cite{Salfi:2014kaa,Sinthiptharakoon:2013il} at $V\approx-0.70$~V, we identify the first electron in a tip-induced QD state when the bands are bent downward for the resonance at $V \approx -1.10$~V, away from the donor.  Notably, the resonance shifts to $V\approx-1.05$~V as the QD approaches the neutral donor showing that the coupled donor/QD state has a lower energy than the isolated donor and QD, since less downward bias of the localized state (relative to the reservoir) is required for resonant tunneling. 

The observed local dip of the QD resonance near a neutral donor in Fig.~\ref{fschem}c is inconsistent with a non-interacting state of the donor and QD where the QD energy would not depend on tip position (Fig.~\ref{fschem}d, lower black line).  To explain the data we need to consider spin singlet paired two-electron states S$(i,j)$ with $i$ electrons on the donor and $j$ on the QD. Charging an isolated donor with a second electron can also be ruled out, since a parabolic S(2,0) resonance would be expected in this case (Fig.~\ref{fschem}d, upper black line). This is because as the tip moves away from the donor it is less effective at locally influencing the potential at the donor site\cite{Teichmann:2008bh} so a larger bias is needed to overcome the donor's on-site Coulomb repulsion. In contrast, the resonance in Fig.~\ref{fschem}c flattens out, approaching S(1,1)-like behaviour. Consequently the two-electron (2e) state in Fig.~\ref{fschem}c can only be understood as a hybridized superposition of S(1,1) and S(2,0) singlets (Fig.~\ref{fschem}d, solid blue). 

\begin{figure}[b]
\includegraphics{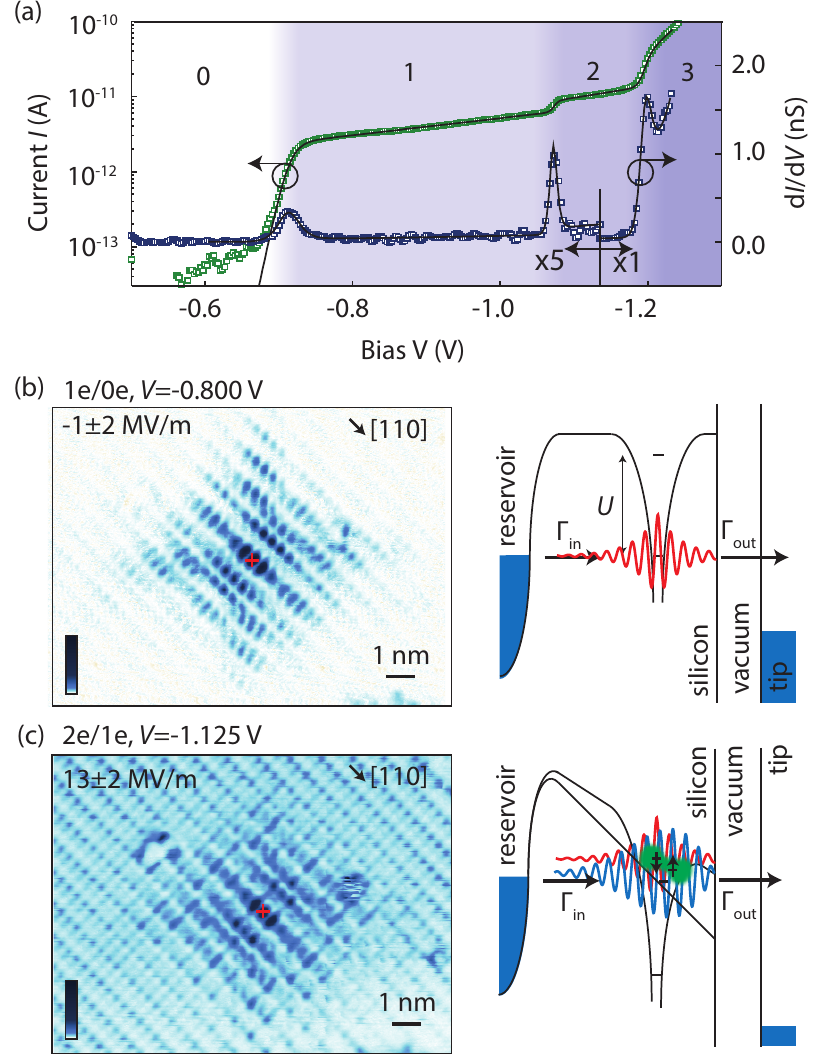}
\caption{(a) Bias dependence of current $I$ (green) and conductance ${\rm d} I/{\rm d} V$ (blue) over the donor, taken at $-1.42$ nm in Fig.~1c, and fit to rate-equation model (black lines). (b) Measured current map of the neutral donor. Right: tunnel junction energy diagram. (c) Measured current map of coupled donor/QD singlet. Right: tunnel junction energy diagram. Data for the donor and donor/QD resonances was acquired in the spatial region of the inset of Fig.~1c.  The donor ion position is marked with a red cross.}
\label{fim}
\end{figure}

Importantly, the donor/QD system forms a molecular state in Fig.~\ref{fschem}c since the donor/QD exchange energy $J$ well exceeds the reservoir tunnel rates, $h(\Gamma_{\rm in}+\Gamma_{\rm out})$. This result is obtained by combining two experimentally established inequalities: First ${\rm d} I/ {\rm d} V$ has a lineshape of a thermally broadened reservoir (Fig.~\ref{fim}a), so $k_BT > h(\Gamma_{\rm in}+\Gamma_{\rm out})$. Second, $J$ well exceeds $k_{\rm B}T$, as will be shown later.

Using the spatially resolved map of ${\rm 1e}\rightarrow{\rm 0e}$ tunneling from the donor at $V=-0.80$ V (Fig.~\ref{fim}b) we identify the donor ground state\cite{Miwa:2013ib,Sinthiptharakoon:2013il} by the A$_1$ valley interference pattern\cite{Salfi:2014kaa}. We determine the donor depth $6.75a_0$ beneath the silicon surface ($a_0=0.543$~nm) using a comprehensive tunnelling image analysis\cite{Usman:2016dka}. We assume zero electric field for comparison with Fig.~\ref{fim}b, which is justified since the tip bias $V=-0.80$ V induces a small electric field $-1\pm 2$ MV/m\cite{SMref}.  With the $2{\rm e}$ donor/QD state in the bias window at $V=-1.125$ V, the electric field in the sample is considerably larger ($13\pm1~\text{MV/m}$, see~\onlinecite{SMref}).  With the second resonance in the bias window, spatially resolved electron tunnelling to the tip (Fig.~\ref{fim}c) represents a $2{\rm e}\rightarrow 1{\rm e}$ quasi-particle wavefunction (QPWF)\cite{Rontani:2005eb,Salfi:2016cm,Maruccio:2007kl,Schulz:2015hv}. 

For the discussion of the measured two electron donor/QD hybrid QPWF resonance in Fig.~\ref{fim}c we use the spin singlet model illustrated in Fig.~\ref{fschem}d with $\left|{\rm S}\right\rangle=c_{1,1}\left|{\rm S}(1,1)\right\rangle + c_{2,0}\left|{\rm S}(2,0)\right\rangle$, where $c_{i,j}$ is the probability amplitude for $\left|{\rm S}(i,j)\right\rangle$. For the ${\rm 2e}\rightarrow{\rm 1e}$ transition, the tunneling current is $I(\mathbf{r})=|{\mathbb{D}} (2^{-1/2}c_{1,1}\psi_{\rm d_1}(\mathbf{r}))|^2 + |\mathbb {D}(2^{-1/2}c_{1,1}u_{\rm q}(\mathbf{r})+c_{2,0}\psi_{\rm d_{\rm 2}}(\mathbf{r}))|^2$\cite{SMref}.  Here $\psi_{\rm d_i}(\mathbf{r})$ is the donor orbital for electron number $i$, $\mathbf{r}$ is the movable QD's position relative to the donor, $u_{\rm q}(\mathbf{r})$ is the lattice-scale structure of the moving QD's wavefunction, and $\mathbb{D}$ is a differential operator that takes the STM tip orbital into account\cite{Chen:1990cr}.  We have found exceptionally good agreement of our single donor measurements\cite{Salfi:2014kaa,Voisin:2015gl} with sp$^3$d$^5$s$^\ast$ theory, including $d$-orbital tips\cite{Usman:2016dka}. 

As expected, the centre of the donor/QD QPWF map (Fig.~\ref{fim}c) strongly resembles the measured neutral donor (Fig.~\ref{fim}b) because both S$(2,0)$ and S$(1,1)$ contain donor bound orbitals, as reflected in the above expression for $I(\mathbf{r})$. We note that when the ${2\rm e}\rightarrow {1\rm e}$ transition is in the bias window, the $1{\rm e}\rightarrow 0{\rm e}$ transition also remains energetically allowed (Fig.~\ref{fim}c).  However, following the $2{\rm e}\rightarrow 1{\rm e}$ transition, the $1{\rm e}\rightarrow 0{\rm e}$ transition is much less likely than a $1{\rm e}\rightarrow 2{\rm e}$ transition because the electron loading rate from the reservoir $\Gamma_{\rm in}=\Gamma_{1{\rm e}\rightarrow 2{\rm e}}$ far exceeds the tunnel rate to the tip $\Gamma_{\rm out}=\Gamma_{1{\rm e}\rightarrow 0{\rm e}}$\cite{SMref}.  Hence, the strong appearance of the donor in the QD resonance of Fig.~\ref{fim}c is not due to a ${\rm 1e}\rightarrow {\rm 0e}$ transition.  Rather, it confirms the pairing interaction of the QD with the donor.  

Away from the donor, the donor/QD resonance (Fig.~\ref{fim}c) is lattice periodic in the $(x,y)$ plane as expected for a QD wavefunction containing only $+z$ and $-z$ valleys\cite{Ando:1982jy,Boykin:2004ih}. Importantly, the QD and donor states are expected to have a significant vertical overlap as illustrated in Fig~\ref{fschem}a, since the QD charge density is expected to peak at $z_0 \approx 3.5a_0$, just $\sim 2$ nm from the donor ion at $6.75a_0$. Here, $z_0$ was estimated using the triangular well approximation\cite{Ando:1982jy} with $z_0 \approx 1.32(\hbar^2/2me\mathcal{E}_z)^{1/3}$, where $\mathcal{E}_z=13\pm 2$~MV/m is the electric field and $m$ is the longitudinal electron effective mass in Si\cite{SMref}. 

\begin{figure}
\includegraphics{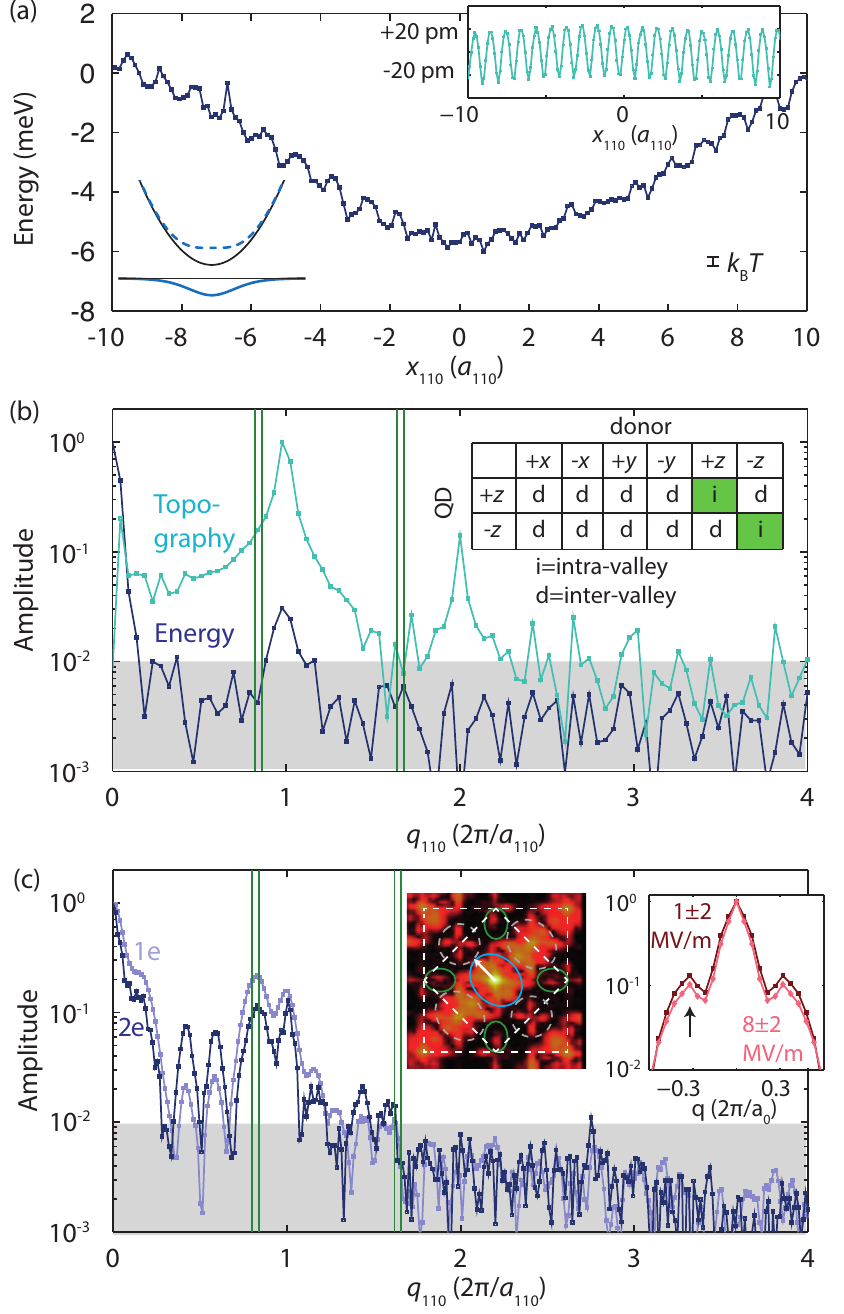}
\caption{a. Donor/QD energy $E(x')$ as a function of tip position $x$. Inset: measured topography. b. Spatial Fourier decomposition of $E(x')$ where lattice-aperiodic components are less than 1~\% of value at $q_{110}=0$. Inset: intravalley (green) and intervalley (white) tunneling for donor and QD. c. One-dimensional spatial Fourier decomposition of 2e and 1e tunnelling current along a $[110]$ direction, with lattice aperiodic oscillations due to valley superpositions in donor bound states at $q_{110}=0.85 (2\pi/a_{110})$ and harmonics, where $a_{110}=\sqrt{2}a_0$. Left inset: two-dimensional Fourier decomposition of one-electron tunneling probability at electric field $8\pm 2$~MV/m and bias $V=-1.010$~V. Right inset: line cut through two-dimensional decomposition evidencing negligible valley repopulation of donor just below the 2e resonance.}
\label{fex}
\end{figure}

The remainder of the analysis focuses on the dependence of the envelope and spatial oscillations present in the donor/QD resonance energy $E(x')$. Plotted in Fig.~\ref{fex}a, $E(x')$ obtained from Fig.~\ref{fschem}c and the extracted lever arm\cite{SMref}, varies by $5.5$ meV over a $16$ nm range of tip positions. Notably, the interface experienced by the QD is atomically flat (Fig.~\ref{fschem}c, inset), which is important since roughness disturbs the valley phase and exchange coupling of QDs\cite{Goswami:2006bb,Nielsen:2012wt,Yang:2013if,Zimmerman:2017hg,Boross:2016cx,Ferdous:2018dy,Huang:2017ec}. The donor/QD energy map $E(x')$ is dominated by exchange coupling $J(x') = J_{\rm dq}(x')+\tfrac{1}{2}(U_{\rm e1}-\sqrt{8t(x')^2+U^2_{\rm e1}})$, which in turn contains tunneling and exchange terms $t(x')$ and $J_{dq}(x')$\cite{Koiller:2001gwa}.  Here, $U_{\rm e1}=E_{\rm d}-E_{\rm q}+U_{\rm dd}-U_{\rm dq}$ is an effective charging energy, $U_{\rm dd}$ is the donor charging energy and $U_{\rm dq}$ is the QD/donor electron repulsion. The remaining contribution to $E(x')$ is the Coulomb interaction of the QD with the neutral donor, estimated to be $\sim 1$ meV\cite{SMref}. Importantly, the Fourier decomposition of $E(x')$  (Fig.~\ref{fex}b) contains no lattice-aperiodic components above $\approx1$~\% of the average of $E(x')$ (at $q_{110}=0$).  Given that $J(x')$ comprises more than 50~\% of $E(x')$ and the residual Coulomb interactions in $E(x')$ do not have lattice aperiodic components, the 1\% upper bound for $E(x')$ corresponds to a 2~\% upper bound of lattice aperiodic components of $J(x')$.  

Tunneling and exchange in a coupled donor/QD system differs from two donors\cite{Cullis:1970jp,Koiller:2001gwa,Wellard:2005hs,Pica:2014hr,Gamble:2015cl,Wang:2016ha} because the QD is a superposition of $\pm z$ valleys only, and the donor is a six-valley superposition as evidenced by lattice-aperiodic components in the Fourier transforms of STM tunnel current maps (Fig.~\ref{fex}c). Notably, the six-valley superposition of the $6.75a_0$ deep donor is hardly affected at all by the QD electrostatic potential, even at an applied electric field of $8\pm 2$ MV/m below the 2e resonance ($V=-1.010$ V). This is evidenced by the Fourier decomposition of the donor measurements below the 2e resonance (Fig.~\ref{fex}c, left inset).  The amplitude of the Fourier peak at ${\bf q}=\pm(+0.15,-0.15)(2\pi/a_0)$ (black arrow in the right inset of Fig.~\ref{fex}c) reflects the $x$ and $y$ valley population\cite{Salfi:2014kaa,Saraiva:2016bs}, and depends very little on the tip bias $V$ in our experiment (Fig.~\ref{fex}c).  By extending our theory comparison\cite{Usman:2016dka} to include electric fields, we estimate a 0.5~\% change in the population of the $x$ and $y$ valleys with the increase in electric field from $-1 \pm 2$ MV/m to $8\pm2$~MV/m due to the STM tip voltage\cite{SMref}.  This is important because for shallower donors where ion-implant statistical uncertainty are suppressed\cite{vanDonkelaar:2015cq}, hybrid donor/QD systems can be formed with negligible perturbation to donor valley composition and hyperfine coupling. 

We note that $E(x')$ contains a lattice periodic oscillation (Fig.~\ref{fex}a, and Fig.~\ref{fex}b). This oscillation is likely an artefact from QD energy and wavefunction changes induced by tip-height variation $\delta z$ (Fig.~\ref{fex}a, inset). For the former, an energy shift of $\delta E = e\alpha \delta z(\epsilon_{\rm Si}\mathcal{E}_z$) is expected\cite{Voisin:2015gl}, where $\alpha\approx0.1$ is the lever arm from our fit. For $\delta z=40$~pm (Fig.~\ref{fex}a, inset) $\delta E=0.6$ meV, in agreement with the measured $0.7$~meV oscillation. Notably, the smooth exchange variation in Fig.~\ref{fex}a indicates that the valley phase\cite{Zimmerman:2017hg,Boross:2016cx,Ferdous:2018dy} varies little, even though the electric field varies by $\sim 3$ MV/m due to the change in the resonance voltage from $V=-1.05$~V to $V=-1.10$~V.  

We now consider lattice-aperiodic oscillations due to interference of valley degrees of freedom in coupled donor/QD systems. The spectral decomposition of $J(\mathbf{R})$ can be theoretically understood from an extended Hubbard model for donor/QD tunneling, $t(\mathbf{R})=\left\langle \psi_{\rm q} | v_{\rm q}+v_{d} | \psi_{\rm d}\right\rangle$, and exchange, $J_{\rm dq}(\mathbf{R})=\left\langle \psi_{\rm d} \psi_{\rm q} | e^2(4\pi\epsilon|\mathbf{r}_1-\mathbf{r}_2|)^{-1} | \psi_{\rm q} \psi_{\rm d}\right\rangle$.  Here, $\mathbf{R}$ is the donor/QD separation, $v_{\rm d(q)}$ is the donor (QD) potential, $\psi_{\rm d}(\mathbf{r})=\sum_{\mu} \psi_{\rm d\mu}(\mathbf{r})$ ($\psi_{\rm q}(\mathbf{r})=\sum_{\mu=\pm z} \psi_{\rm q\mu}(\mathbf{r})$) is the six valley donor (two-valley QD) wavefunction.  In $J(\mathbf{R})$, intravalley (valley preserving) and intervalley (valley modifying) terms have lattice-aperiodic prefactors $\exp(i\mathbf{k} \cdot \mathbf{R})$, since $\mathbf{k}$ values are distributed about the conduction band minima. While the intravalley tunneling present in inter-donor exchange can be evaluated readily\cite{Koiller:2001gwa}, it is expected to be absent here for $x$ and $y$ valleys since they are not present in the QD state. The remaining intervalley processes where electrons change valley index while tunneling\cite{Ning:1971kg,Pantelides:1974ewa} contribute lattice-aperiodic terms $\exp(i\mathbf{k} \cdot \mathbf{R})$ to $J(\mathbf{R})$.  Hence, the $2~\%$ bound on the lattice aperiodic exchange reflects both the effectiveness of valley filtering and provides an upper bound on the strength of intervalley exchange compared to total exchange, which to our knowledge, has not been reported to date. In particular, the intervalley exchange falls outside the scope of the effective mass approximation, but is expected to be enhanced for localized states compared to extended states. This is relevant because of the $\text{nm}$-spatial extent of the localized wavefunctions measured here, which is similar to silicon's lattice constant $a_0=0.543$ nm\cite{Luttinger:1955ee}. 

The experimentally confirmed weakness of the intervalley tunneling means that lateral and vertical donor positioning uncertainty of donors will influence donor/QD exchange in different ways.  Lateral donor positioning uncertainty will influence coupling predominantly through the $\text{nm}$ scale envelope decay length of the QD.  In contrast, vertical donor positioning uncertainty will influence coupling through a combination of the vertical decay length of the donor and QD and interference processes in the intravalley exchange.  The interference should contain an oscillatory term in donor depth because the surface pins the valleys of the QD, while the ion pins the valleys of donor bound electron. The strategy that stands out to compensate these exchange variations is to adjust the QD confinement potential and therefore overlap of the QD state with the fixed donor. This is already accomplished in our experiment since the QD follows the potential of the STM tip, and in proposed devices could be realized by tuning surface gate voltages\cite{Pica:2016bb,Tosi:2017fs}.  In particular, our measurements (Fig.~\ref{fex}a) show that a change of donor/QD separation by $6$ nm changes donor/QD exchange by an order of magnitude, showing that strongly confined QDs allow for a tremendous exchange tuning range. 

To determine if this tuning range is sufficient to overcome intravalley oscillations in exchange due to depth variations in the donors, a quantitative theory analysis has been carried out with atomistic sp$^3$d$^5$s$^\ast$ tight binding.  Experimentally measuring these oscillations is difficult since it would require to ability to change the valley phase of the QD wavefunction, or directly measuring $< 0.5$ meV values of exchange with direct transport, which is not possible in our scheme at $4.2~\text{K}$ .  The QD state in the calculation was calibrated so that full configuration interaction (FCI) wavefunctions\cite{Wang:2016ha,Tankasala:2018by} reproduce experimentally measured spectra. A $5$ nm STM tip radius was found to reproduce the bias where $0{\rm e} \rightarrow 1{\rm e}$ and $1{\rm e} \rightarrow 2{\rm e}$ QD transitions occur, away from the donor. The lowering of the addition energy due to donor/QD coupling when the QD is directly over the donor is calculated to be $E(x'=0)=6.8$ meV, compared to the value $6$ meV in experiments (Fig.~\ref{fex}a).  

\begin{figure}
\includegraphics{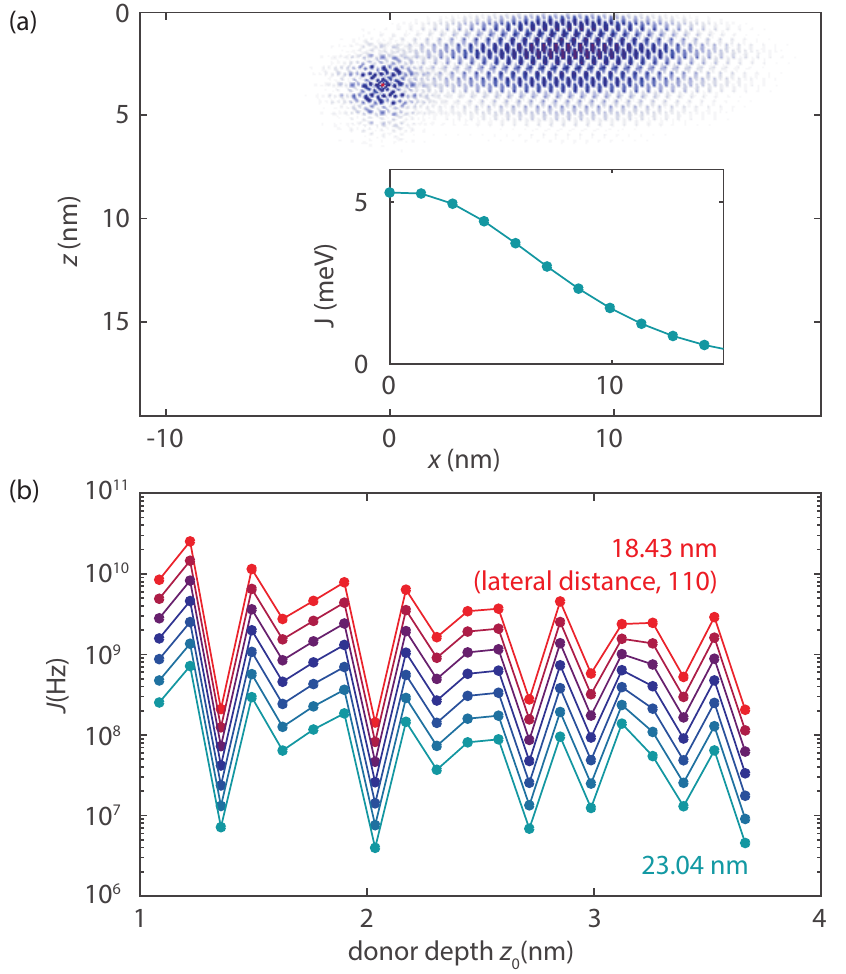}
\caption{a. Calculated electron density for donor and quantum dot, where only the donor contains lattice incommensurate components. Inset: predicted decay of exchange as a function of donor/QD distance.  b.  Calculated dependence of donor/QD exchange on donor depth, for different displacements along 110 between $18.43~\text{nm}$ and $23.04~\text{nm}$.}
\label{fcalc}
\end{figure}

The expected variation in exchange with vertical donor positioning uncertainty was estimated by computing the donor/QD exchange for a range of donor depths and QD distances.  The geometry used in the calculation includes the tip potential and donor ion potential.  A cross section of the calculated charge density including the donor ($x=0$) and QD ($x=8~\text{nm}$) is shown in Fig.~\ref{fcalc}a. In agreement with our measurements, the exchange varies slowly with lateral QD position (Fig.~\ref{fcalc}a, Inset), showing that tight binding accurately reproduces the weak intervalley scattering observed in experiments.   The calculated exchange varies rapidly with donor depth $z_0$ (Fig.~\ref{fcalc}b), but notably, the total variation including the rapidly varying intravalley interference and envelope decay is less than two orders of magnitude for depths between $2.2$ and $3.7$ nm.  This is important because it indicates that variations in exchange due to donor depth uncertainty can be compensated by adjusting donor/QD wavefunction overlap using gates.  The calculations for different donor/QD lateral displacements along the 110 direction (Fig.~\ref{fcalc}b) show that a change of QD position of $\sim 4.5$ nm, between 18.43 nm and 23.04 nm, is sufficient to overcome this variation. We also note that similar to inter-donor exchange, residual coupling uncertainty can in principle be corrected by quantum control\cite{Testolin:2007ku,Hill:2007ku}. 

In conclusion, we have spatially mapped the energy of a neutral donor coupled to a single-electron QD that can be positioned in the plane with sub-nm accuracy using an STM tip.  Besides additional applications of coupling to optically active impurities\cite{Koehl:2011fv,Castelletto:2013dg,Brenneis:2015dq,Bourgeois:2015ke,Buckley:2017hu,Beaufils:2018fa} or small-scale quantum simulators\cite{Salfi:2016cm,Le:2017jt,Dusko:2018gv}, our results highlight that, similar to predictions for donor/donor interactions in strained Si, donor/QD interactions \cite{Srinivasa:2015bm,Pica:2016bb,Tosi:2017fs,HarveyCollard:2017ic} do not suffer from valley-induced variations in exchange due to in-plane donor positioning uncertainty. The demonstrated monotonic tunability of donor/QD exchange with QD position is therefore promising for the realisation of uniform exchange couplings between highly coherent donors using tunable, electrostatically defined QDs\cite{Srinivasa:2015bm,Pica:2016bb,Tosi:2017fs}, that are compatible with an all donor based approach\cite{Kane:1998ce,Hill:2015fj}. 

\begin{acknowledgements} The authors would like to thank D. Culcer for helpful discussions. We acknowledge support from the ARC Centre of Excellence for Quantum Computation and Communication Technology (CE110001027), and partial support from the US Army Research Office (W911NF-08-1-0527). JS acknowledges  support from an ARC DECRA fellowship (DE160101490). The authors acknowledge the use of computational resources from NanoHUB.org/NCN, and the Pawsey Supercomputing Centre with funding from the Australian Government and the Government of Western Australia.  This work used the Extreme Science and Engineering Discovery Environment (XSEDE) ECS150001, which is supported by National Science Foundation grant number ACI-1548562\cite{Towns2014abc}.
\end{acknowledgements}

\section*{Supplemental Material}
\subsection{Electric field experienced by donor and QD}
\label{sEfield}
\begin{figure}[b]
\includegraphics{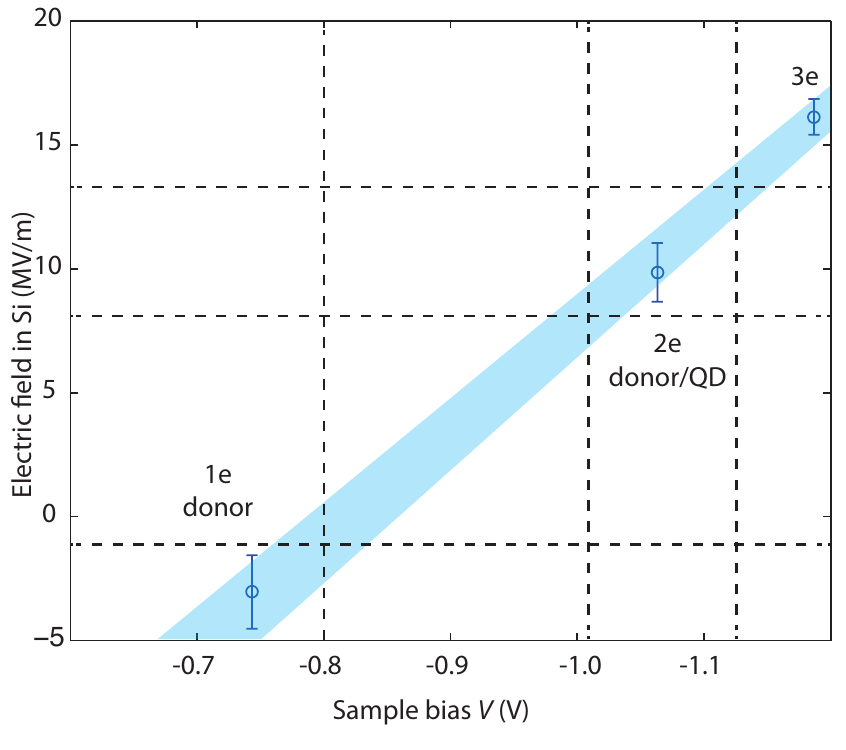}
\caption{Bias dependence of electric field extracted from effective one-dimensional model for tip-sample junction.}
\label{fEfield}
\end{figure}

In this section we experimentally extract the vertical component $\mathcal{E}_z$ of the tip-induced electric field, which eventually confines a single electron against the vacuum interface of Si in the undoped layer of the sample. It has been modelled using a simple one-dimensional electrostatic description accounting for the dependence of the single-electron tunneling peak voltage $U_i$ on tip height $z$.  In this approach, the electric field in the vacuum is given by $-dU_i/dz$, which is reduced by silicon's relative dielectric constant $\epsilon_{\rm Si}$ giving $\mathcal{E}_z=-(\epsilon_{\rm Si}^{-1})dU_i/dz$\cite{Salfi:2014kaa,Voisin:2015gl}.  Electric fields extracted using this procedure are plotted against bias in Fig.~\ref{fEfield}. Three different peak voltages from Fig.~1c in the main text were tracked: the donor 1e resonance ($i=1$), the donor/QD 2e resonance ($i=2$), and the 3e resonance ($i=3$). 
\subsection{Donor depth and repopulation of valleys due to electric fields in neutral donor state}
\begin{figure}
\includegraphics{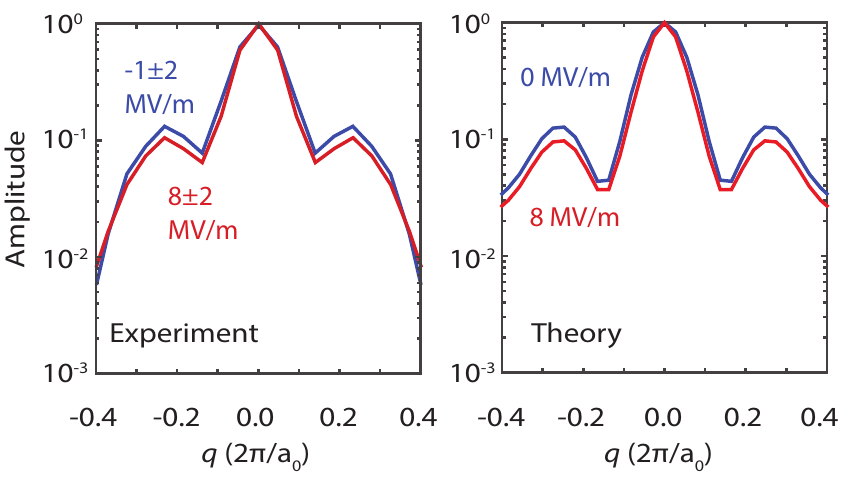}
\caption{Profile of donor state Fourier decomposition along $[110]$ direction for (a) measured STM tunnel current and (b) theoretically predicted STM tunnel current, for two different electric fields.}
\label{fValley}
\end{figure}

In this section we perform and theory/experiment comparison showing that the electric field of the tip has a small impact on the valley population one-electron ground state of the donor. As a baseline, we start with wavefunction measurements in the smallest electric field (Fig.~\ref{fEfield}, $V=-0.80$~V, $\mathcal{E}=-1\pm 2$~MV/m), and where single dopants can be pinpointed with lattice precision in three dimensions\cite{Usman:2016dka}. We re-plot the Fourier representation of the tunnel current along a $[110]$ direction for $\mathbf{q}$, for the $6.75a_0$ deep donor discussed throughout the main text ($a_0=0.543$ nm) shown in Fig.~\ref{fValley}a in the main text.  Plotted alongside this data is the same quantity for an applied electric field $\mathcal{E}_z\approx 8\pm 2$ MV/m at a sample bias $V=-1.010$ V, the same data is shown in Fig.~\ref{fschem}c of the main text. Here, the ratio of the side peak at $q\approx 0.22 (2\pi/a_0)$ and the main peak at $q=0$, given in Table~\ref{tValley}, is proportional to product of $x$ and $y$ valley population\cite{Salfi:2014kaa}. 

The Fourier decomposition of the predicted STM image for this depth is shown in Fig.~\ref{fValley}b, for a uniform electric field $\mathcal{E}_z=0$ MV/m and 8 MV/m, showing a good agreement with experiment with the tip-induced potential.  A detailed analysis of the valley population of the donor, carried out on the tight-binding wavefunction using a basis change with the Slater orbitals described elsewhere\cite{Salfi:2014kaa}, reveals only a 1\% change in $z$ valley population for the higher field.  Based on the good theory/experiment match and the valley repopulation estimate for the theory donor, we estimate the change in $z$ valley population of the measured donor, due to the tip-induced potential that confines the quantum dot (QD), is only around $1$~\%. 

\begin{table}
\begin{tabular} {| l | l | l | l |}
\hline
Electric Field & Expmt/Theory & 0 MV/m & 8 MV/m\\
\hline
\hline
Side-lobe ratio & Expmt. & 0.13 & 0.10\\
\hline
Side-lobe ratio & Theory & 0.13 & 0.09\\
\hline
Valley population $x,y$ (\%)  & Theory & 32.8 & 32.3\\ 
\hline
Valley population $z$ (\%) & Theory & 34.5 & 35.4\\
\hline
\end{tabular}
\caption{Comparison of measured and calculated Fourier spectrum for $z_0=6.75a_0$ deep donor in two different electric fields.}
\label{tValley}
\end{table} 
\subsection{Single-electron tunneling}
In the following analysis we show how to extract addition energies for the 1e, 2e, and 3e transitions.  We also show that in the limit $\Gamma_{\rm in} \ll \Gamma_{\rm out}$, the presence of the $2\rightarrow 1$ charge transition in the bias window blocks the $1\rightarrow 0$ tunneling from contributing to the total current.  We employ a classical rate equation analysis of electron tunneling to describe single-electron transport through multiple charge levels in our donor/QD system\cite{Bonet:2002da} with thermally broadened reservoirs.  The starting point for this model is in ref.~\onlinecite{Voisin:2015gl}.  Here, this model is generalized to non-zero temperature and multiple charge states.  Within this framework, the total current is given by
\begin{equation}
I=\sum_{i} e\rho_{i}\left(\Gamma^{t}_{i\rightarrow i-1}\right)
\end{equation}
Here, $\rho_{i}$ is the probability of $i$ electron occupation.  Moreover, $\Gamma^{t}_{i \rightarrow i\pm 1}$ is the tunnel rate from (to) the tip causing a charge transition $i\rightarrow i\pm 1$ on the donor/QD system, which is given by the product of a bare tunnel rate $\Gamma^t_{i,i\pm 1}$ and a Fermi-Dirac statistical distribution $f(E_F,E)=(1+\exp((E-E_F)/k_BT))^{-1}$ for the tip, as follows:
\begin{align}
\Gamma^{t}_{i\rightarrow i-1}&=\Gamma_{i,i-1}^t[1-f(eV,E_{i}-e\alpha_i V)]\\
\Gamma^{t}_{i-1\rightarrow i}&=\Gamma_{i,i-1}^tf(eV,E_{i}-e\alpha_i V)\\
\end{align}
\begin{figure}
\includegraphics{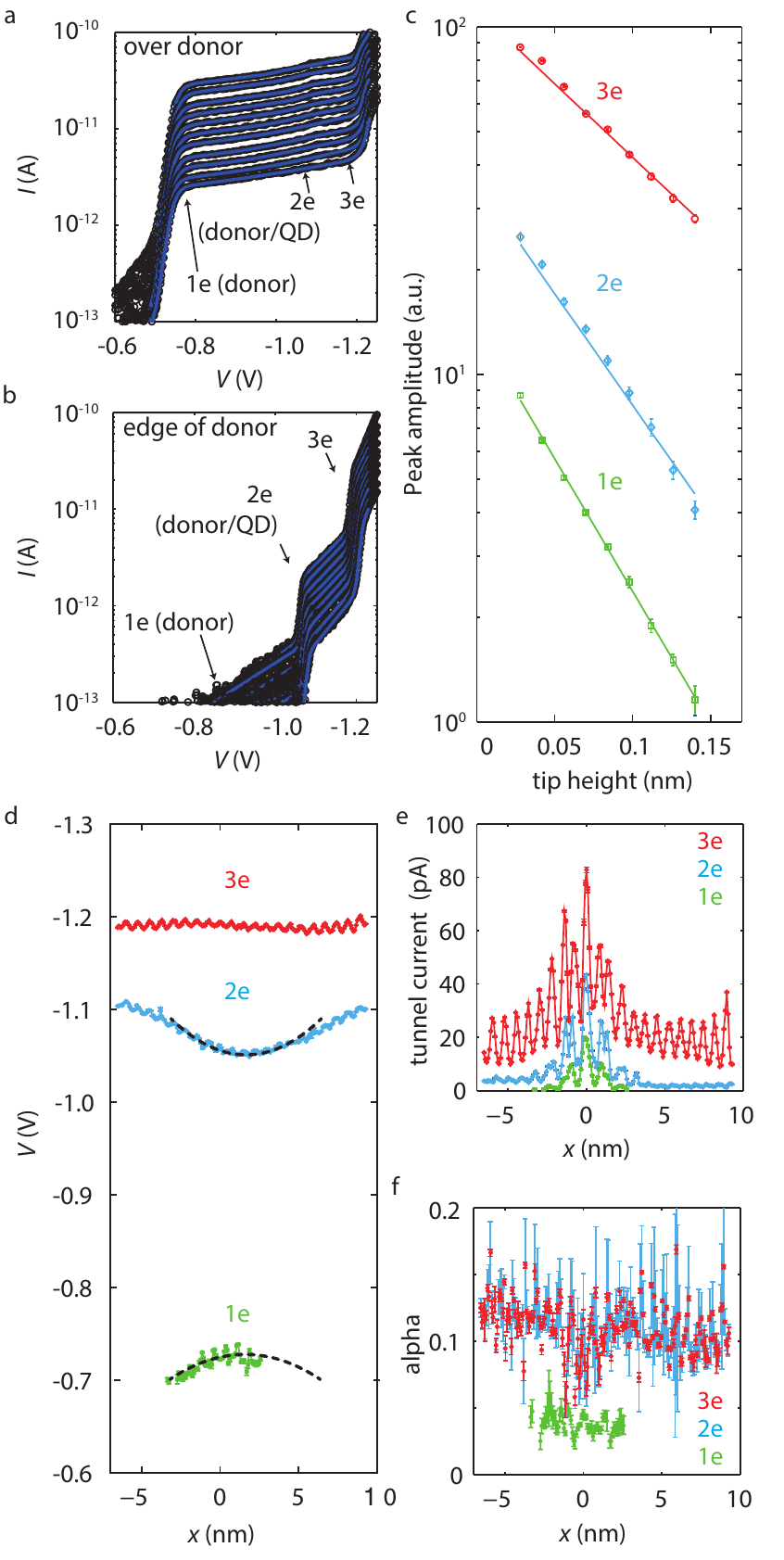}
\caption{Current-voltage characteristic of donor/QD system, for a sequence of fixed equally spaced tip heights (a) for tip position at brightest central dimer in donor and (b) for tip position a few nm away from brightest central dimer of donor.  c.  Dependence of tunnel current amplitude $M^{t0}_{i,i-1}$ step as a function of tip height, and charge number $p$ in the transition $i\rightarrow i-1$. Spatial fits of single-electron tunneling quantities for the 1e, 2e, and 3e resonances, where $x=0$ denotes the position of the donor ion.  (d) transition voltage $V_i$, (e) transition current amplitude $e\Gamma^{t_0}_{i,i-1}$ (f) tip-sample lever arm $\alpha_i$.  The dashed parabolic lines in (d) show that the 1e and 2e resonances are centred along the same x coordinate, as expected for tip-induced band bending\cite{Teichmann:2008bh}. }
\label{fSingleElectron}
\end{figure}
where $\alpha_i$ is the lever-arm for the charge state $i$ with bias $V$, $k_B$ is Boltzmann's constant, $T=4.2$ K is the sample temperature, and $E_{i}$ is the addition energy of the $i$-electron state.  Note that in our experiment, the reservoir chemical potential is well below the states being probed, relative to temperature, so that electron tunneling from the tip to the bound state $\Gamma^{t}_{i-1\rightarrow i}$ is negligible.

We use a master equation to solve for $\rho_i$ given in ref.~\onlinecite{Bonet:2002da}, which also contains tunneling in and out of the reservoir in the sample, to the quantized states.  Noting that the current depends exponentially on tip height, (Fig.~\ref{fSingleElectron}a,b), we can conclude that $\Gamma_{\rm out}=\Gamma^t_{i,i-1}$ to the tip is much less than the $\Gamma_{\rm in}=\Gamma^r_{i-1,i}$ from the reservoir.  To zeroth order in $\Gamma_{\rm out}/\Gamma_{\rm in}$, our rate equation model yields
\begin{equation}
I = \sum_i \Gamma^t_{i,i-1}f_i(1-f_{i+1})
\label{eqIre}
\end{equation}
where $f_i=(1+\exp(-e\alpha_i(V-V_i)/k_{\rm B}T))^{-1}$ is the probability of an electron in the sample reservoir at the energy $E_i=e\alpha V_i$ for the $i^{\rm th}$ charge state. 

We fit the measured $z$-dependent spectrum in Fig.~\ref{fSingleElectron}a,b.  We employ an exponential barrier lowering with bias described by $\gamma_i$ in $\Gamma^t_{p,p\pm 1}=\Gamma^{t_0}_{p,p\pm 1}\exp(\gamma_i(V-V_i))$.  Fits for the tunnel current in the centre of the donor, and a few nm away from the donor, are shown for different tip heights $z$ as blue lines superimposed on the data in Fig.~\ref{fSingleElectron}a and Fig.~\ref{fSingleElectron}b, and are in excellent agreement with the data.  The extracted values for $\Gamma^{t_0}_{i,i-1}$ as a function of tip height are shown in Fig.~\ref{fSingleElectron}c, and demonstrate the exponential tip height dependence of current from experiments, as expected.  The solution to the rate equations in the limit of small $\Gamma^t_{i,i-1}/\Gamma^r_{i,i+1}$ given in Equation \ref{eqIre} establishes the result that when the $2\rightarrow 1$ charge transition enters the bias window, the tunnel current reflects only this transition, and blocks the $1\rightarrow 0$ transition from contributing to the total tunneling current. Finally, the extracted energy for the two-electron state $E_i$ is given in Fig.~3a in the main text.

We fit the $x$-dependent spectral data in the main text, Fig.~1c to obtain the voltage $V_i$, lever arm $\alpha_i$ and tunnel rate $\Gamma^t_{i,i-1}$ in this model, and extract the energy $E_i=e\alpha_i V_i$ using $V_i$ and spatially smoothed values of $\alpha$, as plotted in Fig.~\ref{fex}a.  The fit of the peak voltage, current, and lever arms are shown Fig.~\ref{fSingleElectron}d, e and f.  Notably for the peak voltage, close to the donor, the parabolic dependence on tip voltage is centred along the same coordinate $x_{\rm o} \sim 2$ nm.  This offset from the donor center $x=0$ probably reflects the difference in the location of the centre of mass of the QD wavefunction, and the atom position where tunneling to the tip occurs. 

\subsection{Simplified Hubbard model for spectrum and images}

In this section we theoretically discuss a Hubbard model for the energy spectrum and images that is motivated by features observed in the experiments. Here, the donor-QD system is described by potentials $v_{\rm d}$ of the donor and $v_{\rm q}$ of the QD.  Defining two single-electron Hamiltonians $h_{\rm d}=\mathscr{T}+v_{\rm d}$ and $h_{\rm q}=\mathscr{T}+v_{\rm q}$, where $\mathscr{T}$ is the kinetic energy operator, the states of interest in the separated systems corresponding to the donor and QD are $\phi_{\rm d}$ and $\phi_{\rm q}$ respectively, which satisfy equations $h_{\rm d}\phi_{\rm d}=\varepsilon_{\rm d}\phi_{\rm d}$ and $h_{\rm q}\phi_{\rm q}=\varepsilon_{\rm q}\phi_{\rm q}$ respectively. Then in the composite system defined by the total applied potential $v_{\rm d}+v_{\rm q}$, the one-electron problem is determined by the Hamiltonian $h=\mathscr{T}+v_{\rm d}+v_{\rm q}$, and the two-electron problem is determined by the Hamiltonian $h=h^{(1)}+h^{(2)}+V^{(12)}$ where $h^{(1)}=\mathscr{T}^{(1)}+v^{(1)}_{\rm d}+v^{(1)}_{\rm q}$ is a function of coordinate $\mathbf{r}_1$ only, $h^{(2)}=\mathscr{T}^{(2)}+v^{(2)}_{\rm d}+v^{(2)}_{\rm q}$ is a function of coordinate $\mathbf{r}_2$ only, and $V^{(12)}$ is the electron-electron Coulomb repulsion. 

To solve the one and two-electron problems we construct a Wannier basis of maximally localized orbitals\cite{Schliemann:2001iz}, $\psi_{\rm d}=(\phi_{\rm d}-g\phi_{\rm q})/\sqrt{1+g^2-2gS}$ and $\psi_{\rm q}=(\phi_{\rm q}-g\phi_{\rm d})/\sqrt{1+g^2-2gS}$, where $S=\bk{\phi_{\rm d}|\phi_{\rm q}}$ is the overlap.  The value $g=(1-\sqrt{1-S^2})/S$ ensures the orthogonality of the Wannier orbitals, \textit{i.e.}, $\bk{d|q}=\bk{q|d}=0$, while the normalization $\bk{d|d}=\bk{q|q}=1$ is ensured by the prefactor $1/\sqrt{1+g^2-2gS}$.

\subsubsection{One-electron spectrum}
For the one-electron Hamiltonian in the Wannier basis $\{\psi_{\rm d}(\mathbf{r}),\psi_{\rm q}(\mathbf{r}-\mathbf{R})\}$, the eigenstates obey
\begin{equation}
\left(\left(\frac{E_{\rm d}+E_{\rm q}}{2}\right)\mathbb{I}+\left(\begin{array}{cc}
-\frac{\Delta}{2} & t(\mathbf{R})\\
t(\mathbf{R})&\frac{\Delta}{2}
\end{array}\right)\right)\Psi=E\Psi
\end{equation}
where $\mathbb{I}$ is the identity matrix, $\Delta=E_{\rm q}-E_{\rm d}$ is the donor/QD detuning, $t=\bk{{\rm d}|h|{\rm q}}$ is the tunneling from $\psi_{\rm d}$ to $\psi_{\rm q}$ due to the total potential of the donor and tip, and energies $E_{\rm d}=\bk{{\rm d}|h|{\rm d}}=\varepsilon_{\rm d}+\bk{{\rm d}|v_{\rm q}|{\rm d}}$ and $E_{\rm q}=\bk{{\rm q}|h|{\rm q}}=\varepsilon_{\rm q}+\bk{{\rm q}|v_{\rm d}|{\rm q}}$ reflect confinement by $v_{\rm d}$ and $v_{\rm q}$ due to the donor and tip.  The eigenenergies are:
\begin{equation}
E_1=\frac{E_{\rm d}+E_{\rm q}}{2}\pm\frac{1}{2}\sqrt{\Delta^2+4t^2}
\end{equation}
\subsubsection{Two-electron spectrum}
We expand the two-electron problem in a basis of singlets and triplets of the QD and donor Wannier functions written above.  The singlets $\Psi_{\rm S}$ and triplets $\Psi_{\rm T}$ are 
\begin{widetext}
\begin{align}
&\Psi_{\rm S(1,1)}(\mathbf{r}_1,\mathbf{r}_2)=\tfrac{1}{\sqrt{2}}\left(\psi_{\rm d}(\mathbf{r}_1)\psi_{\rm q}(\mathbf{r}_2-\mathbf{R})+\psi_{\rm q}(\mathbf{r}_1-\mathbf{R})\psi_{\rm d}(\mathbf{r}_2)\right)\ket{\rm S}\\
&\Psi_{\rm S(2,0)}(\mathbf{r}_1,\mathbf{r}_2)=\psi_{\rm d}(\mathbf{r}_1)\psi_{\rm d}(\mathbf{r}_2)\ket{\rm S}\\
&\Psi_{\rm S(0,2)}(\mathbf{r}_1,\mathbf{r}_2)=\psi_{\rm q}(\mathbf{r}_1-\mathbf{R})\psi_{\rm q}(\mathbf{r}_2-\mathbf{R})\ket{\rm S}\\
&\Psi_{\rm T^0}(\mathbf{r}_1,\mathbf{r}_2)=\tfrac{1}{\sqrt{2}}\left(\psi_{\rm d}(\mathbf{r}_1)\psi_{\rm q}(\mathbf{r}_2-\mathbf{R})-\psi_{\rm q}(\mathbf{r}_1-\mathbf{R})\psi_{\rm d}(\mathbf{r}_2)\right)\ket{\rm T^0}\\
&\Psi_{\rm T^-}(\mathbf{r}_1,\mathbf{r}_2)=\tfrac{1}{\sqrt{2}}\left(\psi_{\rm d}(\mathbf{r}_1)\psi_{\rm q}(\mathbf{r}_2-\mathbf{R})-\psi_{\rm q}(\mathbf{r}_1-\mathbf{R})\psi_{\rm d}(\mathbf{r}_2)\right)\ket{\rm T^-}\\
&\Psi_{\rm T^+}(\mathbf{r}_1,\mathbf{r}_2)=\tfrac{1}{\sqrt{2}}\left(\psi_{\rm d}(\mathbf{r}_1)\psi_{\rm q}(\mathbf{r}_2-\mathbf{R})-\psi_{\rm q}(\mathbf{r}_1-\mathbf{R})\psi_{\rm d}(\mathbf{r}_2)\right)\ket{\rm T^+}
\end{align}
\end{widetext}
where $\ket{\rm S}=\frac{1}{\sqrt{2}}\left(\ket{\uparrow\downarrow}-\ket{\downarrow\uparrow}\right)$, $\ket{\rm T^0}=\frac{1}{\sqrt{2}}\left(\ket{\uparrow\downarrow}+\ket{\downarrow\uparrow}\right)$, $\ket{\rm T^+}=\ket{\uparrow\uparrow}$, and $\ket{\rm T^-}=\ket{\downarrow\downarrow}$.

We evaluate all the matrix elements of $h=h^{(1)}+h^{(2)}+V^{(12)}$ in the above basis. The singlet and triplet subspaces separate due to their spin orthogonality.  The singlet subspace is described by
\begin{widetext}
\begin{equation}
\left((E_{\rm d}+E_{\rm q}+U_{\rm dq}+J_{\rm dq}(\mathbf{R}))\mathbb{I}+\left(\begin{array}{ccc}
0         & \sqrt{2}t(\mathbf{R}) & \sqrt{2}t(\mathbf{R})\\
\sqrt{2}t(\mathbf{R}) & U_{\rm e1} & 0\\
\sqrt{2}t(\mathbf{R}) & 0      & U_{\rm e2}
\end{array}\right)\right)\Psi=E\Psi
\end{equation}
\end{widetext}
where $U_{\rm e1} = (E_{\rm d}+U_{\rm dd})-(E_{\rm q}+U_{\rm dq}+J_{\rm dq}(\mathbf{R}))$ is an effective charging energy to put two electrons on the donor, $U_{\rm e2} = (E_{\rm q}+U_{\rm qq})-(E_{\rm d}+U_{\rm dq}+J_{\rm dq}(\mathbf{R}))$ is an effective charging energy to put two electrons on the QD, $U_{\rm qq}=\bk{\rm dd|V_{12}|\rm dd}$ is the charging energy of the donor level, $U_{\rm dq}=\bk{{\rm dq}|V_{12}|{\rm dq}}$ is the mutual Coulomb repulsion of the QD and donor, $U_{\rm qq}=\bk{{\rm qq}|V_{12}|{\rm qq}}$ is the charging energy of the QD level, and $J_{\rm dq}=\bk{{\rm qd}|V_{12}|{\rm dq}}$ is an exchange interaction.  The simplest limit is to diagonalize the upper $2\times2$ block assuming that double occupation of the QD is unlikely ($U_{\rm e2}-U_{\rm e1} \gg t$), giving
\begin{equation}
E_S=E_{\rm d}+E_{\rm q}+U_{\rm dq}+J_{\rm dq}+\tfrac{1}{2}\left(U_{\rm e1}-\sqrt{8t^2+U_{\rm e1}^2}\right),
\end{equation}
while the triplet energy is given straightforwardly by $E_T=E_{\rm d}+E_{\rm q}+U_{\rm dq}-J_{\rm dq}$.  

\begin{figure}
\includegraphics{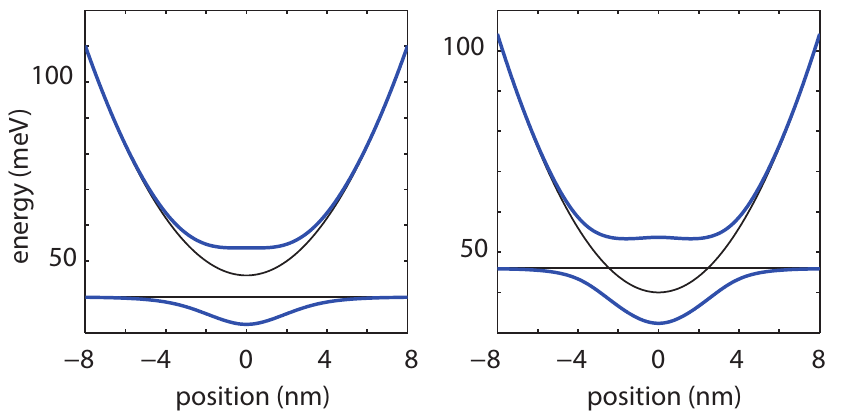}
\caption{A. Case where $U_{\rm e1}>0$ (black lines) and with tunnel coupling turned on (blue lines).  B. Case where $U_{\rm e1}<0$ (black lines) and with tunnel coupling turned on (blue lines).}
\label{fLevels}
\end{figure}

Two cases, $U_{\rm e1}>0$ (Fig.~\ref{fLevels}a, blue lines) and $U_{\rm e1}<0$ (Fig.~\ref{fLevels}b, blue lines) give qualitatively similar results for $E_{\rm S}$. For reference, we show the results ignoring tunnel couplings (black lines, Fig.~\ref{fLevels}a,b).  

\subsubsection{Donor/QD resonance transition energy}

When the single-electron donor/QD detuning $\Delta$ is much larger than the two-electron effective charging energy, as expected for our experiments, we obtain a transition energy
\begin{equation}
E_{\rm S}-E_1=E_{\rm q}+U_{\rm dq}+J_{\rm dq}+\tfrac{1}{2}\left(U_{\rm e1}-\sqrt{8t^2+U^2_{\rm e1}}\right).
\end{equation}
From the definitions of $E_{\rm q}$ and $U_{\rm dq}$ we obtain
\begin{equation}
E_{\rm q}+U_{\rm dq}=\varepsilon_{\rm q}+\bk{{\rm q}|v_{\rm d}|{\rm q}} + U_{\rm dq},
\end{equation}
where $\varepsilon_{\rm q}$ is the energy of the non-interacting QD's ground state, and the final two terms $E_{\rm q, D^0}=\bk{{\rm q}|v_{\rm d}|{\rm q}} + U_{\rm dq}$ are the Coulombic interaction of the QD with the neutral donor. Finally we rewrite the transition energy as
\begin{equation}
E_{\rm S}-E_1=\varepsilon_{\rm q}+E_{\rm q,D^0}+J_{\rm dq}+\tfrac{1}{2}\left(U_{\rm e1}-\sqrt{8t^2+U^2_{\rm e1}}\right).
\end{equation}
The purely Coulombic interaction $E_{\rm q,D^0}$ of the neutral donor and the QD was calculated using
\begin{widetext}
\begin{equation}
E_{\rm q, D^0}=\bk{{\rm q}|v_{\rm d}|{\rm q}} + U_{\rm dq}=\int {\rm d}r_2^3 |\psi_{\rm q}(\mathbf{r}_2-\mathbf{R})|^2 \frac{-e^2}{4\pi\epsilon_0\epsilon_{\rm Si}r_2}+\int {\rm d}r_1^3 {\rm d}r_2^3  |\psi_{\rm q}(\mathbf{r}_2-\mathbf{R})|^2\frac{e^2}{4\pi\epsilon_0\epsilon_{\rm Si}|\mathbf{r}_1-\mathbf{r}_2|}|\psi_{\rm d}(\mathbf{r}_1)|^2
\end{equation}
\end{widetext}
where we have assumed the donor is at the origin. Evaluating this integral using a Monte Carlo technique we find that $E_{\rm q, D^0}$ varies in space and peaks at $\sim 1$ meV for $\mathbf{R}=0$ when the QD overlaps the donor strongest.  The small value of this interaction is attributed to the fact that the donor is neutral. Then, the transition energy should be dominated by exchange terms
\begin{equation}
J(\mathbf{R})=J_{\rm dq}(\mathbf{R})+\tfrac{1}{2}\left(U_{\rm e1}-\sqrt{8t(\mathbf{R})^2+U^2_{\rm e1}}\right)
\end{equation}
accounting for most of the $5.5$ meV spatial variation in energy $E(x)$ for Fig.~3a in the main text.  
\subsubsection{$2\rightarrow 1$ transition image}
The donor/QD system forms a molecule weakly probed by single-electron tunneling by the reservoirs\cite{Salfi:2016cm}, such that the STM image represents a quasi-particle wavefunction\cite{Rontani:2005eb}.  For our two-electron state $\Psi_{\rm S}(\mathbf{r}_1,\mathbf{r}_2)=c_{11}\Psi_{\rm S(1,1)}(\mathbf{r}_1,\mathbf{r}_2)+c_{20}\Psi_{\rm S(2,0)}(\mathbf{r}_1,\mathbf{r}_2)$ and $2\rightarrow1$ transition, the quasiparticle wavefunction is
\begin{equation}
\Psi_{\rm Q}^{2\rightarrow1i}(\mathbf{r})=\int d\mathbf{r}'\Psi^*_{\rm S}(\mathbf{r}',\mathbf{r})\psi_i(\mathbf{r}')
\end{equation}
for a single-electron final state $\psi_i(\mathbf{r})$. Then the total current is the sum of currents for each possible final state, given by $I(\mathbf{r})=\sum_i |\mathbb{D}(\Psi_{\rm Q}^{2 \rightarrow 1i}(\mathbf{r}))|^2$, where $\mathbb{D}$ is a derivative operator accounting for the orbital content of the STM tip\cite{Chen:1990cr}.  The tip orbital with d-like symmetry $d_{z^2-1/3r^2}$ has been found to be important to describe real space STM images of donor-bound electrons in silicon\cite{Usman:2016dka}. Considering two possible final states, the one-electron donor state and the one-electron QD state, we obtain
\begin{align}
I(\mathbf{r})=&|\mathbb{D}(c_{11}2^{-1/2}\psi_{\rm q}(\mathbf{r})+c_{20}\psi_{\rm d_2}(\mathbf{r}))|^2+\\\nonumber&|\mathbb{D}(c_{11}2^{-1/2}\psi_{\rm d_1}(\mathbf{r}))|^2.
\end{align}
Since the QD follows the tip, we note that only the lattice-periodic component of the QD wavefunction $u_{\rm q}(\mathbf{r})$ can be detected, so we must replace $\psi_{\rm q}(\mathbf{r})$ with $u_{q}(\mathbf{r})$.  Doing this we obtain the expression for $I(\mathbf{r})$ presented in the main text.

\subsection{Full configuration interaction model}
\begin{figure}
\includegraphics{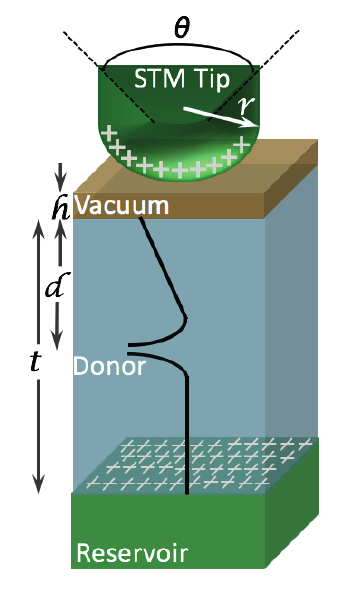}
\caption{Schematic of parameters used to describe the electrostatics of the tip and sample. Here, $\theta$ = the opening angle of tip, $r$ = tip radius, $h$ = tip height above silicon/vacuum interface.  Two additional parameters $d$ describe the donor depth and $t$ the sample reservoir depth below the silicon/vacuum interface. }
\label{fTipSample}
\end{figure}

A full configuration interaction approach using tight-binding wavefunctions, used in ref.~\onlinecite{Tankasala:2018by} to model two-electron states of donors in uniform electric fields, was used to model the interactions of the donor and tip-induced QD. Here, the uniform electric field is replaced by a non-uniform potential of an STM tip, as necessary for the STM tip to induce a QD as observed.  We describe a procedure for calibrating parameters for the tip-induced potential shown schematically in Fig.~\ref{fTipSample}, and give some details on how the results quoted in the main text are obtained. Note that the depth $d$ is fixed by single-donor metrology, for measurements taken near the flat-band condition\cite{Usman:2016dka}.

First, we performed electrostatic calculations of the STM/vacuum/silicon junction using finite element analysis. Then we obtain the single electron energies and wavefunctions of the dot or hybrid donor-dot using atomistic the tight-binding technique including the electrostatic potential. The simulations were done in NEMO3D\cite{Rahman:2009cc}. Then using the single electron wavefunctions, a basis of two-electron Slater Determinants is constructed and full configuration interaction calculations are performed taking into account the image charges, as described in ref.~\onlinecite{Tankasala:2018by}. Diagonalizing the FCI Hamiltonian, the computed 2e total energies, charging energies and singlet-triplet splittings were obtained.

The STM tip parameters were fixed by comparing the binding energy and charging energy of the QD state when it is far away from the donor, at the boundaries of Fig.~1c in the main text.  In our experiment, relative to the flat-band voltage ($V\approx -0.8$ V, Fig.~\ref{fEfield}), a sample bias $\delta V=-0.3$ V (actual bias $V=-1.1$ V) and $\delta V=-0.4$ V (actual bias $V=-1.2$ V) are required to bring the state into resonance with the sample reservoir.  We could reproduce these binding and charging transitions using a tip radius of $r=5$ nm, for an expected reservoir depth of $t=15$ nm, a tip opening angle assumed to be $\theta=45$ $^\circ$ and tip height $h=0.15$ nm.  Variations in the tip opening angle and tip height were found to have less of an influence than the tip radius and reservoir depth.  Reproducing the binding and charging transition voltages gives us confidence that the Bohr radius of the tip-induced QD is similar to the actual value in experiments.  

For these tip and reservoir parameters, we estimated the modulation of the two-electron energy when the single-electron QD interacts with the single electron of the neutral donor at the depth determined from our experiments.  In our transport experiment this is equivalent to comparing the $\rm{0e} \rightarrow \rm{1e}$ transition energy of the QD state in the absence of interactions with the donor (the QD binding energy), to the $\rm{1e} \rightarrow \rm{2e}$ transition energy of the QD/donor state, when the tip-induced dot is directly above the donor.  

For a donor at $6.75a_0$ we obtain $\Delta E$ = 53.3 meV - 46.5 meV = 6.8 meV, which is very similar to the value 5.5 meV from experiments.  Assuming a donor one unit cell closer to the surface at $5.75a_0$ gives $\Delta E$ = 56.7 meV - 46.5 meV = 10.2 meV, a larger interaction.  This is because a donor closer to the surface has a larger overlap, and therefore exchange interaction, with the QD. 

merlin.mbs apsrev4-1.bst 2010-07-25 4.21a (PWD, AO, DPC) hacked
Control: key (0)
Control: author (0) dotless jnrlst
Control: editor formatted (1) identically to author
Control: production of article title (0) allowed
Control: page (1) range
Control: year (0) verbatim
Control: production of eprint (0) enabled

\end{document}